\documentclass[pra,letterpaper,aps,floatfix,superscriptaddress,twocolumn]{revtex4}
\usepackage{graphicx}
\usepackage{amsmath,amssymb}
\usepackage{bm}

\usepackage[pdfstartview=FitH,breaklinks=true,bookmarks=true,colorlinks=true,anchorcolor=black,citecolor=red,filecolor=black,menucolor=black,urlcolor=blue,linkcolor=blue]{hyperref}

\begin{document}

\title{Interaction-induced Metal to Topological Insulator Transition}

\author{Yu-Chin Tzeng}
\affiliation{Physics Division, National Center for Theoretical Sciences, Taipei 10617, Taiwan}
\affiliation{Department of Electrophysics, National Yang Ming Chiao Tung University, Hsinchu 300093, Taiwan}
\author{Po-Yao Chang}
\affiliation{Department of Physics, National Tsing Hua University, Hsinchu 30013, Taiwan}
\author{Min-Fong Yang}
\email{mfyang@thu.edu.tw}
\affiliation{Department of Applied Physics, Tunghai University, Taichung 40704, Taiwan}

\date{\today}

\begin{abstract}
%
By means of exact diagonalizations, the Bernevig-Hughes-Zhang model at quarter-filling in the limit of strong Hubbard on-site repulsion is investigated. We find that the non-interacting metallic state will be turned into a Chern insulator with saturated magnetization under strong correlations. That is, at such a metal-insulator transition, both the topological and the magnetic properties of the system are changed due to spontaneous breaking of time reversal symmetry in the ground states. According to our findings, this topological phase transition seems to be of first order.
Our results illustrate the interesting physics in topological Mott transitions and provide guidance to the search of more interaction-induced topological phases in similar systems.
\end{abstract}

\maketitle

\section{introduction}

The exploration of phase transitions driven by electron correlations is one of the main themes of modern condensed matter physics. A classic paradigm is the Mott transition, which occurs when a metallic system becomes an insulator owing to strong electronic interactions~\cite{Mott1974,Imada_etal1998}.
The basic idea comes from that, with interactions, singly and doubly occupied states are no longer energetically degenerate. Once the doubly occupied states were all pushed up to high energy, the half-filled metallic system would be turned into an insulating one and the gap opens entirely due to the interactions.
The simplest model describing this transition is the repulsive one-band Hubbard model at half-filling~\cite{Hubbard1963}. Unfortunately, beyond the one-dimensional case, exact solutions are not available and we have to rely upon approximate solutions or numerical simulations. Therefore, many important features of the metal-insulator transitions (MITs) have not been understood completely.

Alternatively, one can tackle the Mott physics by formulating exactly solvable toy models which share some properties with realistic ones. Many interesting properties can thus be simulated in a comprehensive way. To this end, instead of the on-site Hubbard interaction, Hatsugai and Kohmoto considered the one being local in momentum space and then of infinite range in real space~\cite{Hatsugai-Kohmoto1992}. The advantage of this Hatsugai-Kohmoto (HK) model is that all the calculations can be performed analytically. Moreover, this model has the same atomic and band limits as the Hubbard model and describes a MIT. Interestingly, it is claimed recently that  the essence of the Mott transition is captured only by the local-in-momentum part of the Hubbard interaction~\cite{Phillips2022}. As a result, it is sufficient to focus on the HK interaction that breaks the relevant symmetry.

After the discovery of topological insulators~\cite{TI_rev1,TI_rev2}, topological phases of matter have gained much attention in the past decades. Understanding the interplay of topology and electron interactions then becomes at the forefront of current research in condensed matter physics~\cite{Budich-Trauzettel2013,Hohenadler-Assaad2013,Meng_etal2014,Rachel18}. Among such research directions, an intriguing one is to find interaction-induced topological phases, such as the fractional quantum Hall effect~\cite{FQHE1,FQHE2}.
Motivated by the important role of the HK interaction in MIT, the Mott physics driven by this interaction is investigated recently for several models of topological insulators~\cite{Phillips22_1,Phillips22_2}. It is found that the quarter-filled metallic states in both the Kane-Mele (KM)~\cite{KM2005} and the Bernevig-Hughes-Zhang (BHZ)~\cite{BHZ2006} models can be driven to be quantum spin-Hall (QSH) Mott insulators under strong HK interactions~\cite{Phillips22_2}. Interestingly, the strongly correlated phases at quarter-filling are predicted to carry a spin Chern number with half the non-interacting value at half-filling. This conclusion is further supported numerically by the determinantal quantum Monte Carlo (DQMC) calculations with the on-site Hubbard interaction. Similar result is obtained as well for the strongly interacting spinful Haldane model~\cite{Haldane1988}, where a Chern Mott insulator appears at quarter-filling instead~\cite{Phillips22_1}.

While the conclusion in Ref.~\cite{Phillips22_2} are interesting, due to the limitation of the employed methods, the predicted QSH Mott insulators may not be the true ground state at low temperatures. First, the local-in-momentum HK interaction always gives a huge spin degeneracy in the strongly correlated phases, because each momentum state is singly occupied by either spin-up or -down electrons. However, unique ground states with spontaneous symmetry breaking usually occurs at low temperatures for most realistic models. Second, restricted by the sign problem in DQMC, the temperatures in the numerical simulations can not be low enough. Therefore, this approach may fail to detect the true low-temperature phases.

In the present work, the method of exact diagonalizations is employed to determine the strongly correlated ground state at zero temperature. Here we focus on the BHZ model at quarter-filling with the on-site Hubbard interaction, but the same conclusion might apply to the KM model as well. To calculate the (spin) Chern number of the interacting systems, the real-space many-body marker recently proposed in Ref.~\cite{Gilardoni_etal2022} is employed. We find that the topologically non-trivial Mott insulating state at zero temperature is a ferromagnetic Chern Mott insulator, rather than the QSH Mott insulator. That is, the time-reversal symmetry is broken spontaneously and then the ground state carries a nonzero Chern number for the selected spin polarization. In addition, depending on system parameters, the strongly correlated insulating state can be topologically trivial Mott insulators as well. To provide the overall picture of the MITs, the zero-temperature phase diagram is presented. Combining the findings in Ref.~\cite{Phillips22_2} and ours, there may exist a finite-temperature phase transition between the high-temperature spin-unpolarized QSH Mott phase and the low-temperature ferromagnetic Chern Mott insulator. At such finite-temperature transition points, both the magnetic and the topological characters are changed.

This paper is organized as follows. Our model Hamiltonian is introduced in Sec.~II and basic properties of the topological marker proposed in Ref.~\cite{Gilardoni_etal2022} are reviewed. Our numerical results are presented in Sec.~III. They provide clear evidences for the existence of interaction-induced topological MITs. We summarize our work in Sec.~IV.

\section{model Hamiltonian and topological marker}

A paradigmatic non-interacting model of QSH insulators is the BHZ model~\cite{BHZ2006}, defined by the Hamiltonian
\begin{align}\label{eq:BHZ}
H_0 &= \sum_\mathbf{k} \Psi_\mathbf{k}^\dagger \mathcal{H}_0 (\mathbf{k})  \Psi_\mathbf{k} \; , \\
\mathcal{H}_0(\mathbf{k}) &= d_x(\mathbf{k})\tau_x\sigma_z + d_y(\mathbf{k})\tau_y\sigma_0 + d_z(\mathbf{k})\tau_z\sigma_0 \; . \nonumber
\end{align}
Here $\Psi_\mathbf{k}=(c_{\mathbf{k}1,\uparrow},\, c_{\mathbf{k}2,\uparrow},\, c_{\mathbf{k}1,\downarrow},\, c_{\mathbf{k}2,\downarrow})^{\rm T}$ with $c_{\mathbf{k}\,l,\sigma}$ being the annihilation operator of a fermion carrying a wave number $\mathbf{k}$, an orbital label $l$, and a spin index $\sigma$. $\tau_a$ and $\sigma_a$ are the Pauli matrices for the orbital and the spin spaces, respectively. For the BHZ model, $d_x(\mathbf{k})=\lambda\sin(k_x)$, $d_y(\mathbf{k})=\lambda\sin(k_y)$, $d_z(\mathbf{k})=m-t\cos(k_x)-t\cos(k_y)$. Here we consider the case of $t=\lambda$ and set $t=1$ as the energy scale. The $z$ projection of the total spin is conserved for the present model and the eigen-energies of the single-particle Hamiltonian $\mathcal{H}_0(\mathbf{k})$ are $\epsilon_{\pm,\mathbf{k}}%
=\pm\sqrt{d_{x}^2(\mathbf{k})+d_{y}^2(\mathbf{k})+d_{z}^2(\mathbf{k})}$ for each spin component. We note that, even with the same energy $\epsilon_{\pm,\mathbf{k}}$, the spin-up and -down electrons have different wave functions with opposite chirality.

At half-filling, the lower bands of both spin components are completely filled. When $0<|m|<2$, the system behaves as a QSH insulator with the Chern number $C\equiv C_{\uparrow}+C_{\downarrow}=0$ and the spin Chern number $C_{s}\equiv C_{\uparrow}-C_{\downarrow}= 2$. In contrast, the system becomes a topologically trivial band insulator with both $C$ and $C_{s}$ being vanishing when $|m|>2$. Below half-filling, the lower bands are partially filled and then the non-interacting system becomes metallic for all values of $m$.

In Ref.~\cite{Phillips22_2}, the authors focus on the $m=1$ case and they show that the non-interacting metallic state at quarter-filling becomes a topological Mott insulator under strong electron interactions. The strongly correlated phase is claimed to be a spin-unpolarized QSH Mott insulator with $C=0$ but $C_{s}=1$. That is, it carries the values with just half the non-interacting ones at half-filling. Their conclusion is led by two independent methods: an exactly solvable model with the HK interaction and the DQMC calculations with the Hubbard interaction. However, as discussed in the previous section, their approaches may fail to predict the true low-temperature phases for realistic interacting systems.

To determine true zero-temperature strongly correlated phases and the phase transitions out of them, we employ the method of exact diagonalizations in the present work.
Following Ref.~\cite{Phillips22_2}, we consider the on-site Hubbard interaction and then the Hamiltonian of the interacting model becomes
\begin{align}\label{eq:model}
H &= H_0 + H_U \; , \\
H_U &= U \sum_i n_{i,\uparrow}\;n_{i,\downarrow} \; . \nonumber
\end{align}
Here $n_{i,\sigma} = \sum_{l=1,2} c_{i\,l,\sigma}^\dagger c_{i\,l,\sigma}$ denotes the total particle number of electrons with spin $\sigma$ on site $i$. In the large $U$ limit, the strong on-site repulsion penalizes the occupation of a site with electrons of different spin. Therefore, spin ordering may emerge in the ground state, if it does not induce further energy cost in kinetic energy.

Depending on the value of $m$, the large-$U$ Mott insulating states could have distinct topological character. To reveal their topological properties, the real-space many-body marker recently proposed in Ref.~\cite{Gilardoni_etal2022} is employed. It is defined by the ratio of three ground-state expectation values,
\begin{equation}\label{eq:rhobhz}
\rho_{\sigma} = \frac{\langle \Psi_0| Z_{\sigma}(\delta k_x,\delta k_y)|\Psi_0\rangle}
{\langle \Psi_0|  Z_{\sigma}(\delta k_x,0)|\Psi_0\rangle \, \langle \Psi_0| Z_{\sigma}(0,\delta k_y)|\Psi_0\rangle} \; .
\end{equation}
Here $|\Psi_0\rangle$ is the many-body ground state of the Hamiltonian $H$ with periodic boundary conditions and $\delta k_x =2\pi/L_x$, $\delta k_y =2\pi/L_y$ for systems on retangular lattices with $N_s=L_x\times L_y$ sites. The operator $Z_{\sigma}(\delta\mathbf{k})$ in this calculation is given by
\begin{equation}\label{eq:hatz}
Z_{\sigma}(\delta\mathbf{k}) = \exp \left(i \delta\mathbf{k}\cdot\sum_j \mathbf{R}_j\,n_{j,\sigma} \right) \; ,
\end{equation}
where $n_{j,\sigma}$ is the spin-projected electron density operator on site $j$ located at the position $\mathbf{R}_j$. This unitary operator represents the lattice position operator of electrons with spin $\sigma$. We note that the many-body marker in Eq.~\eqref{eq:rhobhz} bears some resemblance with the many-body invariant for Chern insulators proposed in Ref.~\cite{Kang_etal2021}.  However, the latter needs calculations for ground states with different boundary conditions and then increases computational effort.

As shown in Ref.~\cite{Gilardoni_etal2022}, the sign of $\rho_{\sigma}$ can be taken as a topological $\mathbb{Z}_2$ marker to determine whether the insulating phases are topological or not. An abrupt jump from negative to positive values of $\rho_{\sigma}$ thus indicates a topological transition. This feature comes from the fact that, for a non-interacting model, the marker behaves as $\rho_{\sigma}\propto\exp\left(i\pi C_{\sigma}\right)$, where $C_{\sigma}$ is the Chern number for spin-$\sigma$ component. Therefore, $\rho_{\sigma}$ is negative in the topological insulating phases with $C_{\sigma}=\pm1$, while it becomes positive in the trivial ones with $C_{\sigma}=0$. This approach has been applied with success to the interacting BHZ model at half-filling even for small lattice sizes~\cite{Gilardoni_etal2022}. Its validity in the present study of Mott insulating phases at quarter-filling is thus expected.

\section{results}

Here, we present our results calculated by using exact diagonalizations for the interacting BHZ model $H$ in Eq.~\eqref{eq:model} on clusters of $N_s$ lattice sites. At quarter-filling, the total electron number $N=N_\uparrow+N_\downarrow$ ($N_\sigma$ denotes the electron number with spin $\sigma$) equals to $N_s$.
Due to the conservation of the $z$ component of total spin $S_z=(N_\uparrow-N_\downarrow)/2$, subspaces of the Hilbert space with different $S_z$ are decoupled. We thus diagonalize $H$ for each subspace with a fixed $S_z$ under $N=N_s$.
We note that the numerical calculations become demanding when $N_s>10$, because there are both spin and orbital degrees of freedom and the local Hilbert space dimension at each site is 16. The translational symmetry thus needs to be imposed for our largest cluster size $N_s=4\times3$. Nevertheless, the largest dimension has still about $1.5\times10^9$ in each momentum sector for $S_z=0$, which requires about 70 gigabytes of computer memory in the two-vector Lanczos algorithm to obtain the lowest eigenvalue.

\begin{figure}[tp]
\includegraphics[width=0.45\textwidth]{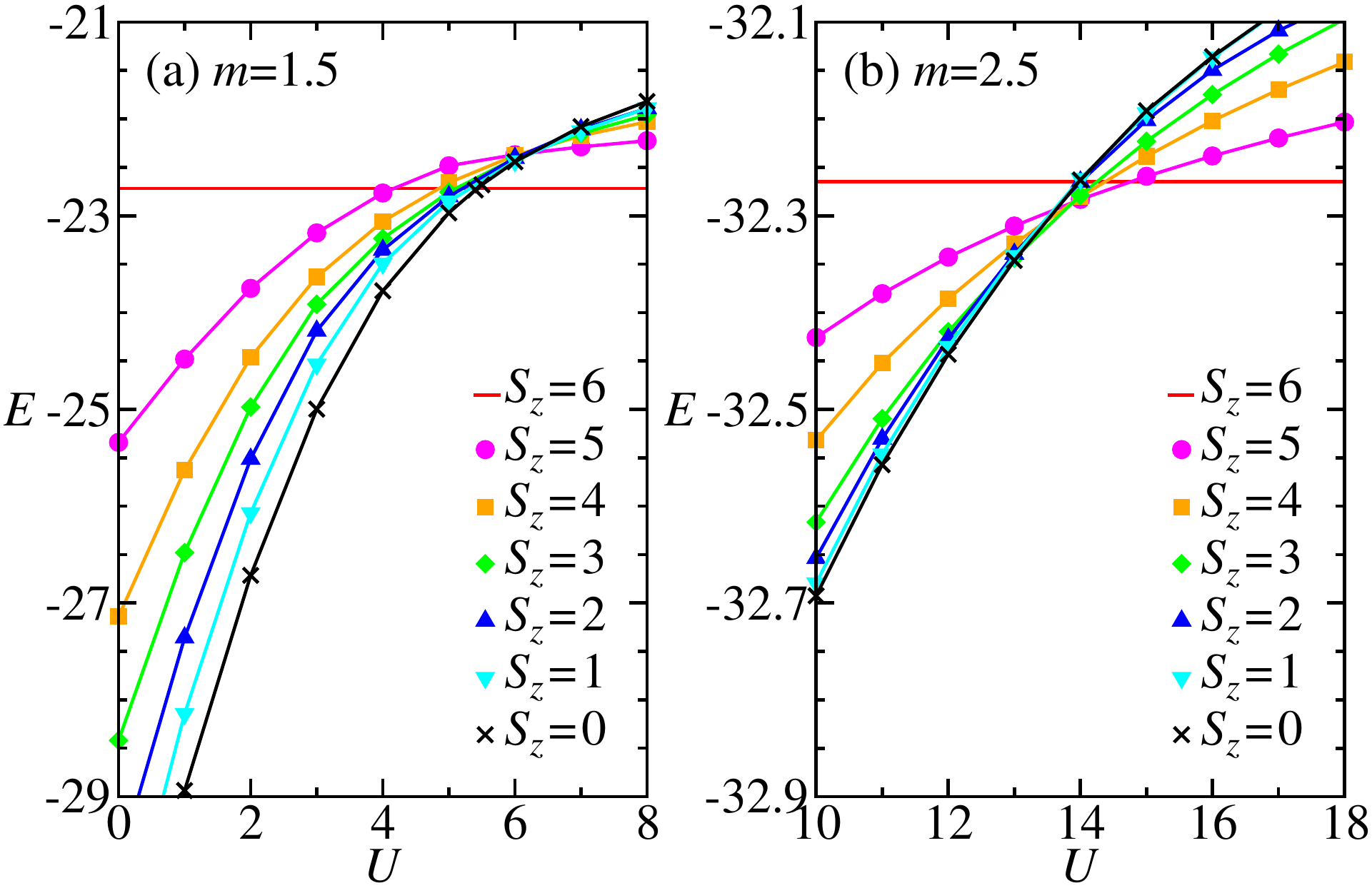}
\caption{%
The lowest energies $E$ as functions of the Hubbard interaction $U$ within each subspace of total $z$-component spin $S_z$ for (a) $m=1.5$ and (b) $m=2.5$. Here the cluster size $N_s=4\times3$ and the total electron number $N=N_s$.
}\label{eng_level}
\end{figure}

In Fig.~\ref{eng_level}, the lowest energies $E$ as functions of the Hubbard interaction $U$ within each subspace are shown for two typical values of $m$. Because of the time reversal symmetry, the results depend only $|S_z|$. Therefore, only the cases of $S_z\geq0$ are presented. The ground state is then given by the one with the minimal energy. We note that the lowest energies $E$ retain their non-interacting values for the fully polarized states with $S_z=N/2$, since the Hubbard repulsion play no role in this case. By contrast, energies of all other states grow as $U$ is increased. The fully polarized states with saturated ferromagnetism thus become the true ground states in the large-$U$ limit for both cases of $m$, while the ground states are paramagnets in the $U=0$ limit.
Our results imply the occurrence of the ferromagnetic transitions in the interacting BHZ model at quarter-filling. When $U$ goes beyond its critical value, one of the two degenerate states with saturated magnetization $S_z=\pm N/2$ is picked up as the ground state such that the time reversal symmetry is broken spontaneously.

Interestingly, these ferromagnetic transitions give the MITs at the same time. The weak-$U$ states with $N_\uparrow=N_\downarrow=N_s/2$ are metallic, because the spin-degenerate lower bands with $\epsilon_{-,\mathbf{k}}$ are both half-filled. These two lower bands can be mixed and then split into the upper and the lower Hubbard bands under strong correlations. At quarter-filling, the lower Hubbard band is now completely filled and the large-$U$ fully-polarized states we found should be insulators. Therefore, MITs induced by the ferromagnetic transitions happen as $U$ is varied from weak to strong.

We note that the presence of the ferromagnetic insulating phases is consistent with the extension of the Lieb-Schultz-Mattis theorem~\cite{LSM_1961, Oshikawa_2000, Hastings_2004} proposed by H. Watanabe \emph{et al.}~\cite{Watanabe_2015}. For fermionic systems with time-reversal symmetry and crystalline symmetries, they show that a unique gapped ground state without spontaneous symmetry breaking is possible only when the mean electron density $\nu$ per unit cell is an even integer. When $\nu$ is odd, the above statement is violated and leads to three possibilities for the ground state: (i) a gapless state, (ii) a gapped state breaking either time-reversal or translational symmetry, (iii) a symmetric long-range entangled Schr\"{o}dinger's cat state. For the system under consideration, the model is time-reversal symmetric and the mean electron density is odd ($\nu=N/N_s=1$). According to the conclusion in Ref.~\cite{Watanabe_2015}, the unique, symmetry-unbroken QSH Mott insulating state suggested in Ref.~\cite{Phillips22_2} can not be the true ground state. Instead, our ferromagnetic insulating states accompanying with spontaneous breaking of time reversal symmetry should give the correct results.

In our calculations for the largest cluster size $N_s=4\times3$, the transition from a weak-$U$ metallic paramagnetic phase to a strong-$U$ insulating ferromagnetic phase behaves as a direct transition for the cases of $m<2$ (e.g., Fig.~\ref{eng_level}(a) for $m=1.5$). On the other hand, an intermediate phase consisting of partially polarized states seems to exist within a narrow window of $U$ for the $m>2$ cases (e.g., Fig.~\ref{eng_level}(b) for $m=2.5$). However, due to possible finite-size effects for small cluster sizes, it is unclear if the intermediate phase would disappear and a single direct transition thus came out in the thermodynamic limit.

Besides the dependence on the interaction $U$, quantum phases can exhibit distinct topological properties for different values of the system parameter $m$. We note that the strongly correlated insulators in the present system are described by the states with the non-interacting lower band $\epsilon_{-,\mathbf{k}}$ being completely filled by spin-up electrons. That is, they are just the spin-up sector of the non-interacting states at half-filling. Nevertheless, their stability at quarter-filling is guaranteed by strong Hubbard repulsion. We then expect that the strong-$U$ states behave as Chern insulators carrying the Chern number $C=C_{\uparrow}=1$ for $0<m<2$ and normal insulators with $C=C_{\uparrow}=0$ when $m>2$. A topological phase transition thus occurs at $m=2$.

\begin{figure}[tp]
\includegraphics[width=0.45\textwidth]{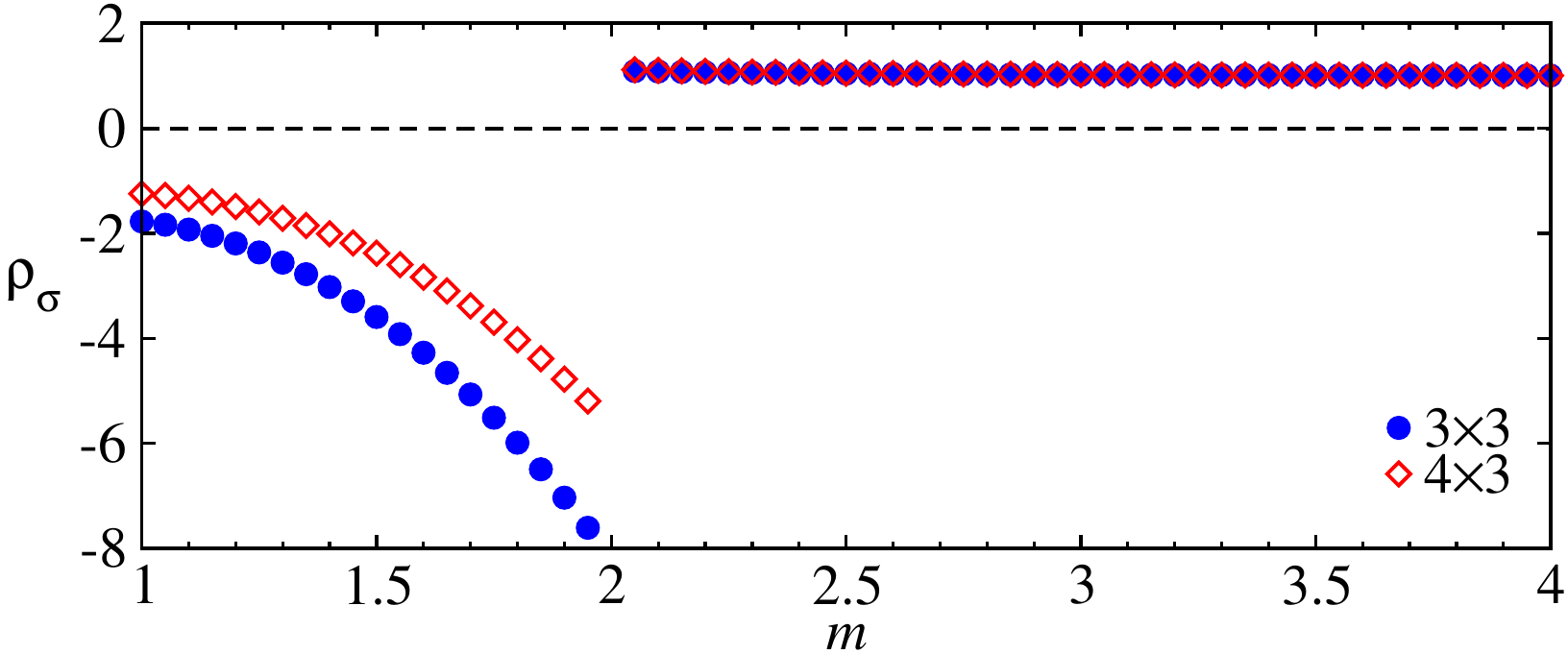}
\caption{%
The many-body $\mathbb{Z}_2$ marker $\rho_{\sigma}$ for spin-up electrons in Eq.~\eqref{eq:rhobhz} calculated by exact diagonalizations for the interacting BHZ model. Here the cluster sizes are $N_s=3\times 3$ and $4\times3$ with $N=N_s$ electrons having parallel spins.
Our findings for $N_s=3\times 3$ are equivalent to those of the $U=0$ case at half-filling shown in Fig.~4 of Ref.~\cite{Gilardoni_etal2022}.
}\label{rhos}
\end{figure}

This conclusion can be checked numerically by using the many-body marker $\rho_{\sigma}$ proposed in Ref.~\cite{Gilardoni_etal2022}. Our findings are shown in Fig.~\ref{rhos}, where the spin component $\sigma$ is taken to be spin-up. As reviewed in Sec.~II, $\rho_{\sigma}$ will be negative in the topological insulating phases with $C_{\sigma}=\pm1$, while it becomes positive in the trivial ones with $C_{\sigma}=0$. Our findings do show negative values for $0<m<2$ and positive ones for $m>2$ with an abrupt change in sign at $m=2$. Therefore, the picture discussed in the last paragraph is verified.

\begin{figure}[tp]
\includegraphics[width=0.45\textwidth]{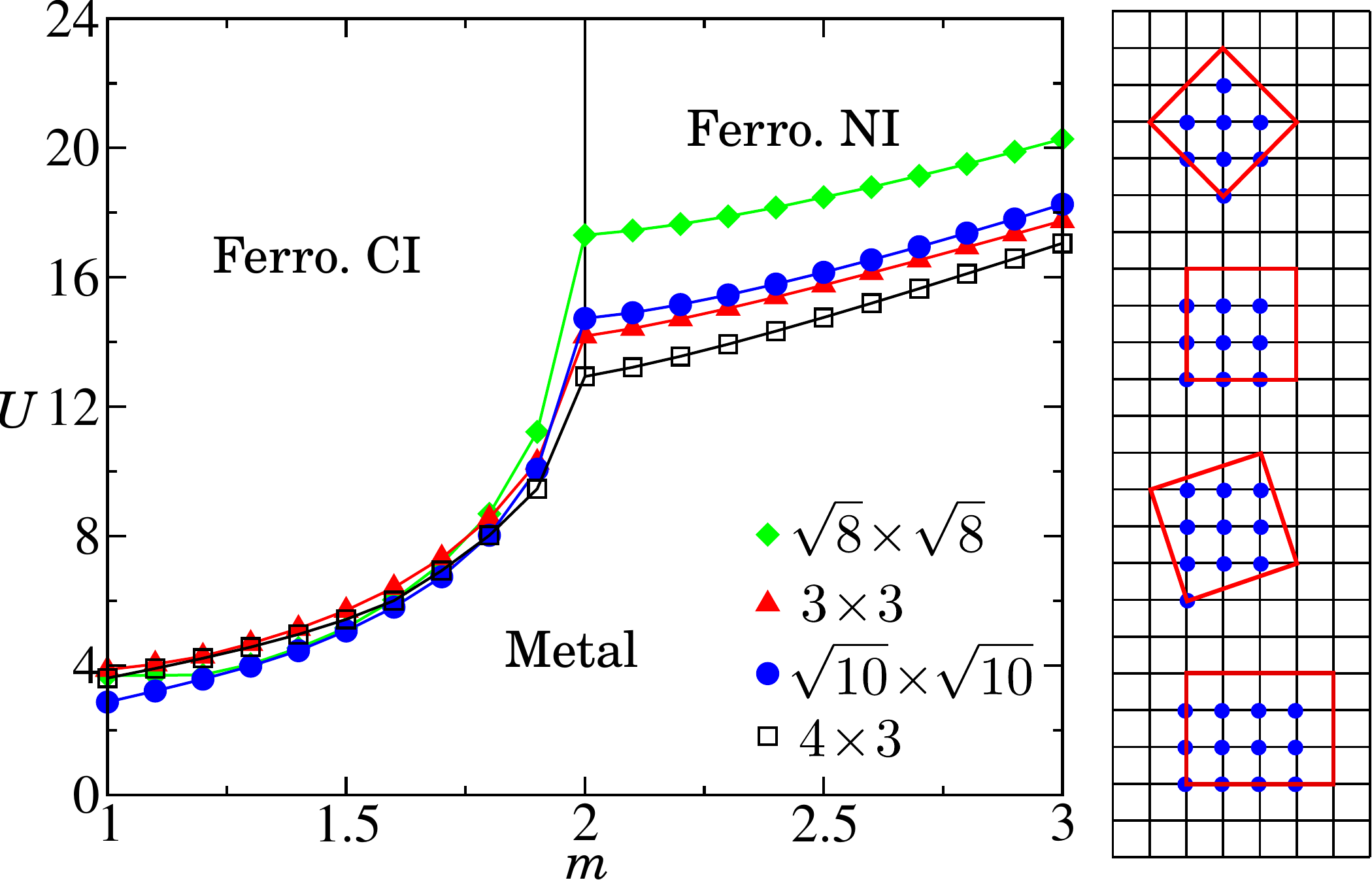}
\caption{%
Quantum phase diagram of the interacting BHZ model in Eq.~\eqref{eq:model} at quarter-filling. The right panel displays the clusters employed in our calculations.
}\label{phase_diagram}
\end{figure}

Our results are summarized by the quantum phase diagram presented in  Fig.~\ref{phase_diagram}. Phase boundaries separating either the ferromagnetic Chern Mott insulators (CIs) or the ferromagnetic normal Mott insulators (NIs) from the metallic phase are determined by the minimal values of $U$ below which the insulating phases become unstable. Their dependence on system sizes is illustrated as well. Because levels of different $S_z$'s cross at those critical $U$'s, the transitions from paramagnetic metals to ferromagnetic insulators (either CIs or NIs) are expected to be of first order. At those MITs, the magnetization jumps from zero to its saturated value. On the other hand, just like the non-interacting cases at half-filing, the topological transitions between the ferromagnetic CIs and NIs at $m=2$ should be continuous.

We note that the critical values of $U$ of the MITs shown in Fig.~\ref{phase_diagram} are always larger than the band widths $W$ of the lower non-interacting band $\epsilon_{-,\mathbf{k}}$. The latter takes the values:
\begin{equation}\label{eq:bandwidth}
W=\left\{
    \begin{array}{ll}
      2m  & \qquad\hbox{for $1\leq m<2$\,;} \\
      4   & \qquad\hbox{for $m\geq2$\,.}
    \end{array}
  \right.
\end{equation}
This implies that perturbation theories and even simple mean-field theory (see App.~\ref{APP} for the analysis of the present model) might not be appropriate to explain the mechanics of the observed transitions.
In our opinions, the saturated ferromagnetism appeared in the strongly correlated phases is reminiscent of the (nearly) flat-band ferromagnetism proposed by Mielke and Tasaki~\cite{Mielke1991,Tasaki1992}, although the energy band under consideration is not flat at all.

Besides, as seen from Fig.~\ref{phase_diagram}, the transitions to Chern Mott insulators always occur at lower critical $U$'s than those to normal Mott insulators. It seems to imply that energy bands carrying non-trivial topological structure could enhance the stability of the insulating phases. Nevertheless, this behavior can be understood qualitatively by analyzing the instability of the ferromagnetic state~\cite{note1}.
In the fully polarized state $|F\rangle$, the lower non-interacting band $\epsilon_{-,\mathbf{k}}$ is completely filled by the spin-up electrons. This state can not be stable once the excitation energy of the single-spin-flip state $|\Psi^\prime\rangle=c^\dag_{\mathbf{p}-,\downarrow}\,c_{\mathbf{q}-,\uparrow}\,%
|F\rangle$ become negative~\cite{note2}. Here the wavevectors $\mathbf{p}$ and $\mathbf{q}$ belong to the band minimum and maximum, respectively. Because $n_{i,\uparrow}\approx1$, the excitation energy $\Delta E$ is about
\begin{equation}
\Delta E = \langle\Psi^\prime|H|\Psi^\prime\rangle - \langle F|H|F\rangle \approx -W + U \; ,
\end{equation}
where the first term is due to the kinetic energy gain and the second one comes from the cost in Hubbard repulsion. This crude estimate shows that the instability of the fully polarized state caused by a single spin flip occurs at $U_c\approx W$. According to the result of the band width $W$ in Eq.~\eqref{eq:bandwidth}, the qualitative behavior of $U_c$ is correctly explained while the value of $U_c$ is underestimated.
Further studies are necessary in order to clarify the complete physics of our discoveries.

\section{conclusions and discussions}

In summary, by using exact diagonalizations, we probe the possible quantum phases in the BHZ model at quarter-filling under the influence of the on-site Hubbard interaction. We find that, at strong couplings, time reversal symmetry can be broken spontaneously and the ground states becomes insulators with saturated ferromagnetism. Our conclusion agrees with the extended Lieb-
Schultz-Mattis theorem~\cite{Watanabe_2015}. It thus lends support to our numerical calculations. Therefore, the existence of interaction-induced MITs accompanied by a ferromagnetic transition is established. In addition, the strongly correlated insulating states can be further classified as either Chern Mott insulators or normal Mott insulators by using their distinct topological properties.

The same model was studied in Ref.~\cite{Phillips22_2} for $m=1$, where a QSH Mott insulator at strong couplings is discovered instead. However, the employed approach is limited to the high-temperature regime and thus possible symmetry broken phases at low temperatures are out of their reach. Therefore, our findings serve as complements to theirs for the low-temperature physics. Integrating the results in Ref.~\cite{Phillips22_2} and ours, there should exist finite-temperature phase transitions between the high-temperature spin-unpolarized QSH Mott phase and the low-temperature ferromagnetic Chern Mott insulator. At such finite-temperature transition points, both the magnetic and the topological characters would be changed. It is interesting to investigate the mechanism and the nature of such finite-temperature topological transitions. The discussions along this direction, however, go beyond the scope of this work.

\begin{acknowledgments}
We are grateful to Federico Becca, Chang-Tse Hsieh, and Shin-Ming Huang for useful discussions.
We acknowledge support from the National Science and Technology Council of Taiwan under Grants No. NSTC 112-2636-M-007-007 and No. MOST 111-2112-M-029-004.
Y.C.T. and P.Y.C. acknowledge the support from the National Center for Theoretical Sciences (NCTS) in Taiwan.
\end{acknowledgments}

\appendix
\section{Mean-field analysis}\label{APP}

\begin{figure*}[tp]
\includegraphics[width=1.\textwidth]{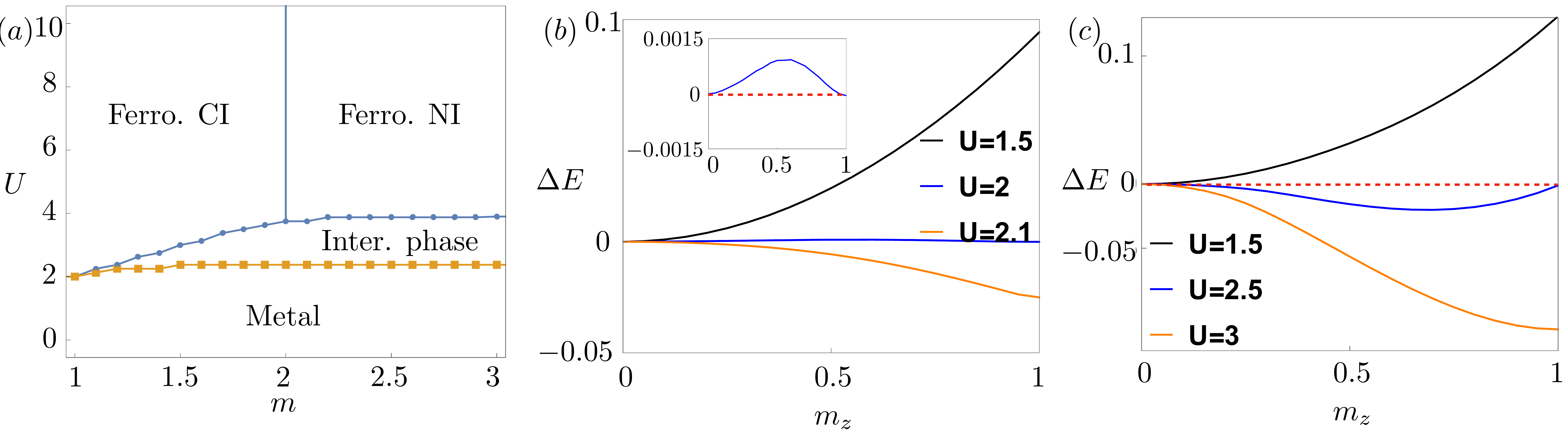}
\caption{ (a) Mean-field phase diagram of the interacting BHZ model in Eq.~\eqref{Eq:MF}.
The differences of energy density $\Delta E=$ versus the magnetization $m_z$ for  (b) $m=1$ and (c) $m=1.5$ and for various $U$ are illustrated. The inset in (b) is the energy density difference $\Delta E$ as a function of $m_z$ for $U=2$ zooming in the region $\Delta E \in [-0.0015,0.0015]$. There are two local minima at $m_z=0$ and $m_z=1$ indicating a first-order phase transition to saturated magnetism.
Here the system size $N_s=101\times 101$ and the total electron number $N=N_s$.
}\label{Fig:MF}
\end{figure*}

In this appendix, we present the mean-field analysis of the BHZ model under the on-site Hubbard repulsion in Eq.~\eqref{eq:model}. Within the present mean-field theory, the upper band $\epsilon_{+,\mathbf{k}}$ becomes irrelevant. Our problem thus reduces to an one-band Hubbard model with energy dispersion of lower band $\epsilon_{-,\mathbf{k}}$ and our treatment becomes nothing but the usual Stoner theory~\cite{Stoner1938,Fazekas}.

From the insight of the exact diagonalizations, the ground state does not break the translational symmetry. Hence we can apply the translation invariant mean-field ansatz to find the mean-field ground state with minimal energy.
We replace the particle number operator $n_{i, \sigma} = \langle n_{\sigma} \rangle + \delta n_{i, \sigma}$ where $\langle n_{\sigma}\rangle$ is the expectation value of the particle number with spin $\sigma$. At quarter-filling, we have $\langle n_{\uparrow}\rangle + \langle n_{\downarrow}\rangle=1$. The corresponding mean-field Hamiltonian is $H_{\rm MF}=\sum_{\bf k}\Psi^\dagger_{\bf k}\mathcal{H}_{\rm MF}({\bf k})\Psi_{\bf k}%
-UN_s(1-m_z^2)/4$ with the single-particle Hamiltonian
\begin{equation}\label{Eq:MF}
\mathcal{H}_{\rm MF}({\bf k})=\mathcal{H}_0 ({\bf k}) + \frac{1}{2}U\tau_0\sigma_0 -\frac{1}{2}U m_z\tau_0\sigma_z \; .
\end{equation}
Here $m_z = \langle n_{\uparrow}\rangle -\langle n_{\downarrow}\rangle$ denotes the magnetization relative to its saturated value.

The single-particle energy is $E_{\alpha,{\bf k}}={\rm Spec} [ \mathcal{H}_{\rm MF} ({\bf k}) ]$ with $\alpha$ being the band label. The lowest many-body energy for each value of $m_z$ is defined as $E(m_z)=\sum_{{\bf k}, \alpha \in {\rm occ.}}  E_{\alpha, {\bf k}}-UN_s(1-m_z^2)/4$ with the single-particle energy being ascending order $ E_{\alpha_1, k_1}< E_{\alpha_2, k_2} < \cdots  E_{\alpha_N, k_N} < \cdots$ and $\sum_{{\bf k}, \alpha \in {\rm occ.}} = N$. The many-body ground state energy $E_G$ is obtained by the lowest many-body energy for various $m_z$. To avoid the size effect and to make a better comparison, we present the difference of the lowest many-body energy density defined as $\Delta E = (E(m_z) - E(m_z=0))/N_s$.

The mean-field phase diagram is shown in Fig.~\ref{Fig:MF} (a).
Unlike the results from the exact diagonalizations, there appears an intermediate phase describing a partially polarization metallic state. As illustrated in Fig.~\ref{Fig:MF}(c) for ($m$, $U$)=(1.5, 2.5), the minimal value of $\Delta E$ occurs at $m_z\simeq0.68$, which indicates a partially polarized state with $m_z \neq 0$ or $1$.

In general, the mean-field analysis may not give a direct first-order phase transition to saturated magnetism as the results obtained by the exact diagonalizations. Nevertheless, such a first-order phase transition does happen at ($m$, $U$)=(1, 2) as shown in Fig.~\ref{Fig:MF}(b). In the inset of Fig.~\ref{Fig:MF}(b), there are two local minima of the energy density for $m_z=0$ and $1$, showing a first-order phase transition.


We finally note that the critical values $U_c$ of the Hubbard interaction that drives the ground state to be a fully polarized insulating state agree with the result $U_c\approx W$ obtained by the analysis of the single-spin-flip instability discussed at the end of Sec.~III. In addition, the predicted $U_c$ follows a similar trend as the results from the exact diagonalizations.



\begin{thebibliography}{76}
\bibitem{note2}
In our case, the true transition should be induced by multiple spin flips, rather than a single one. Nevertheless, the study of the single-spin-flip instabliity may still provide useful insight for the qualtative behavior of the phase boundaries for our MITs.
\end{thebibliography}
\end{document}